\documentclass[journal]{IEEEtran}
\usepackage{graphicx,amssymb,lineno}
\usepackage{amsmath,amsfonts,amssymb}
\usepackage{cases}
\newtheorem{theorem}{Theorem}

\usepackage{algorithm}
\usepackage{algorithmic}
\usepackage[usenames]{color}
\usepackage{subfigure}
\usepackage{graphicx}
\usepackage{subfigure}
\usepackage{setspace}

\title{Exploiting the Shipping Lane Information for Energy-Efficient Maritime Communications }

\author{{Te Wei, \emph{Student Member, IEEE}, Wei Feng, \emph{Member, IEEE}, Jue Wang, \emph{Member, IEEE}, \\ Ning Ge, \emph{Member, IEEE}, Jianhua Lu, \emph{Fellow, IEEE} \vspace{-5 mm}}
\thanks{
T. Wei, W. Feng, N. Ge, and J. Lu are with the Beijing National Research Center for Information Science and Technology, Tsinghua University, Beijing 100084, P. R. China~(e-mail:
kevin.1991@163.com, fengwei@tsinghua.edu.cn, gening@tsinghua.edu.cn, lhh-dee@tsinghua.edu.cn).
J. Wang is with the School of Electronic and Information
Engineering, Nantong University, Nantong 226019, P. R. China.~(e-mail: wangjue@ntu.edu.cn).
}
}

\begin{document}

\maketitle

\begin{abstract}

Energy efficiency is a crucial issue for maritime communications, due to the limitation of geographically available base station sites.
Different from previous studies, we promote the energy efficiency by exploiting the specific characteristics of maritime channels and user mobility.
Particularly, we utilize the shipping lane information to obtain the long-term position information of marine users, from which the large-scale channel state information is estimated.
Based on that, the resource allocation is jointly optimized for all users and all time slots during the voyage, leading to a mixed 0-1 non-convex programming problem. We transform this problem into a convex one by means of variable substitution and time-sharing relaxation, and propose an iterative algorithm to solve it based on the Lagrangian dual decomposition method. Simulation results demonstrate that the proposed scheme can significantly reduce the power consumption compared with existing approaches,
due to the global optimization over a much larger time span by utilizing the shipping lane information.

\end{abstract}


\begin{spacing}{0.95}
\vspace{-0.5 mm}
\section{Introduction}\label{S1}

Unlike terrestrial cellular networks, the maritime communication network (MCN) has to cover a vast area with quite a limited number of geographically available base station (BS) sites.
The network usually adopts high-powered BSs for remote transmission, resulting in low energy efficiency \cite{p333}.
Therefore, it is critically important to design energy-efficient resource allocation strategies for MCNs.

%



Currently, resource allocation techniques have been widely investigated for terrestrial networks.
In \cite{p4}, a joint power allocation and user scheduling algorithm based on dynamic programming was proposed for multi-user multi-input-multi-output (MIMO) systems.
In \cite{p5}, a cross-layer cooperative user scheduling and power allocation scheme was developed for hybrid-delay services.
More recently in \cite{p7}, a user scheduling and pilot assignment scheme was proposed for massive MIMO systems to serve the maximum number of users with guaranteed quality of service (QoS).
These studies identified the importance of acquiring channel state information at the transmitter (CSIT) for enhancing energy efficiency.
Further, to reduce the overhead of acquiring CSIT, the authors in \cite{p52} and \cite{p53} exploited statistical and outdated CSIT for resource allocation.
All of the above
schemes, however, failed to fully utilize the characteristics of user behavior due to the unpredictable user movement.

In contrast to terrestrial networks, user behavior characteristics are exploitable in MCNs,
since most vessels
follow designated shipping lanes.
In \cite{p01}, the authors proposed an opportunistic routing scheme for delay-tolerant MCNs based on lane intersecting opportunities.
In \cite{p8}, the authors proposed three offline scheduling algorithms for video uploading in MCNs based on the deterministic network topology.
These studies utilized the predictability
of user movement, but did not fully
take advantage of
the physical characteristics of maritime channels.
Previous channel modeling studies have suggested that maritime channels consist of only a few strong propagation paths due to the limited number of scatterers, making the slowly time-varying large-scale CSIT more dominant \cite{p2}.
Therefore, acquiring forward-looking large-scale CSIT based on the location information predicted from the shipping lane is more feasible than obtaining instantaneous full CSIT, and is considered as a promising way to improve energy efficiency for
practical
MCNs.

In this paper, we focus on improving energy efficiency for
MCNs based on the shipping lane information. In particular, the resource allocation of all users and all time slots during the service is jointly optimized,
exploiting the large-scale CSIT predicted from the users' position information.
We formulate a long-term optimization problem for the joint allocation of subcarrier and transmit power, aiming to minimize the average power consumption while ensuring the users' QoS requirements.
The problem 
has two non-convex constraints, caused by using large-scale CSIT, and joint optimization for all users and all time slots, respectively. We transform this problem into a convex one by adopting variable substitution and time-sharing relaxation, and propose an iterative algorithm to solve it based on the Lagrangian dual decomposition method.
Simulation results reveal that the proposed forward-looking large-scale CSIT aided scheme significantly reduces the power consumption compared with the schemes using instantaneous full CSIT.

\vspace{-0.5 mm}
\section{System Model} \label{sec:1}

\begin{figure*}
  \centering
  \subfigure[ ]{
    \includegraphics[width=1.712in]{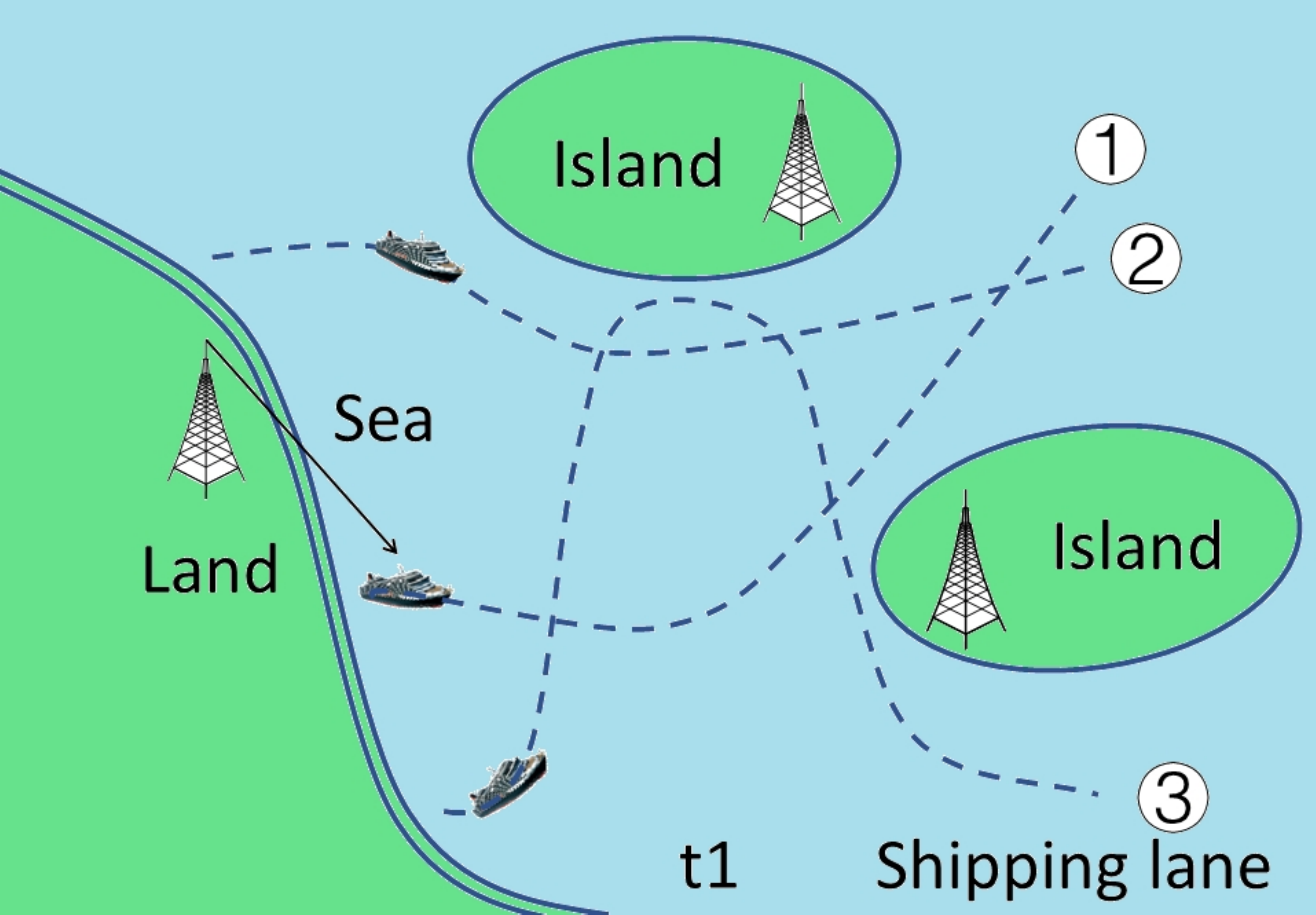}
  }
  \hspace{-3mm}
  \subfigure[ ]{
    \includegraphics[width=1.712in]{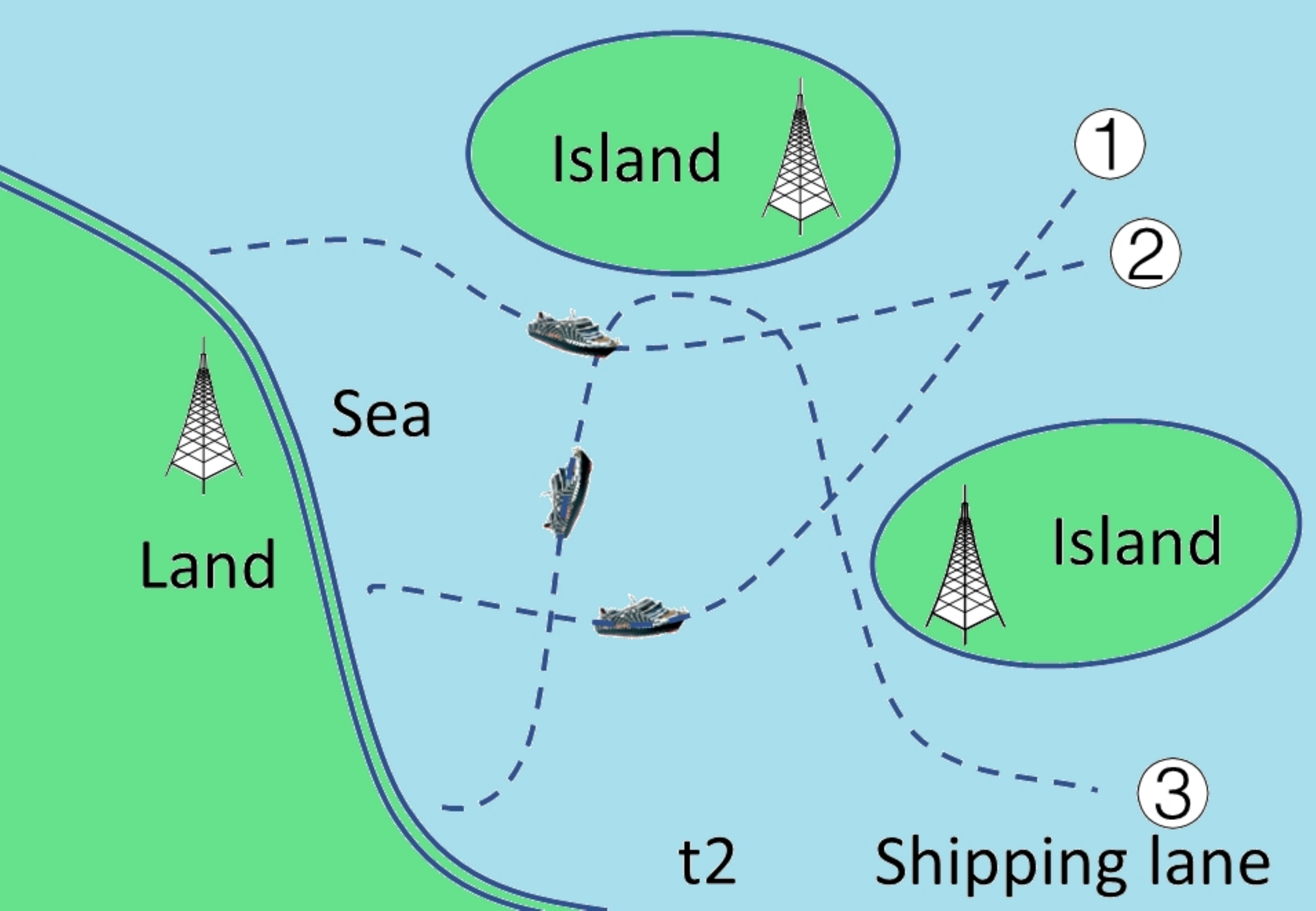}
  }
  \hspace{-3mm}
  \subfigure[ ]{
    \includegraphics[width=1.712in]{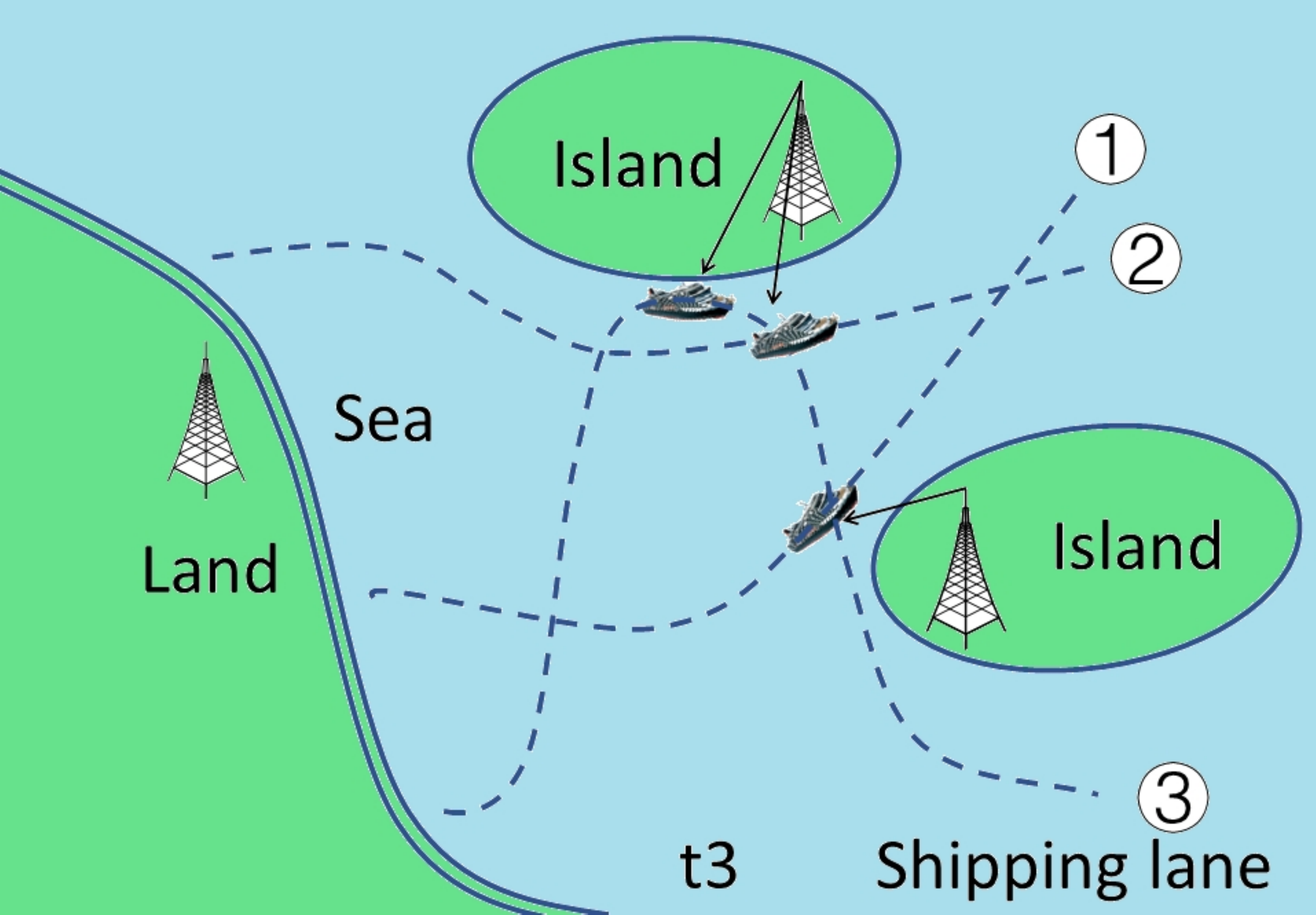}
  }
  \hspace{-3mm}
  \subfigure[ ]{
    \includegraphics[width=1.712in]{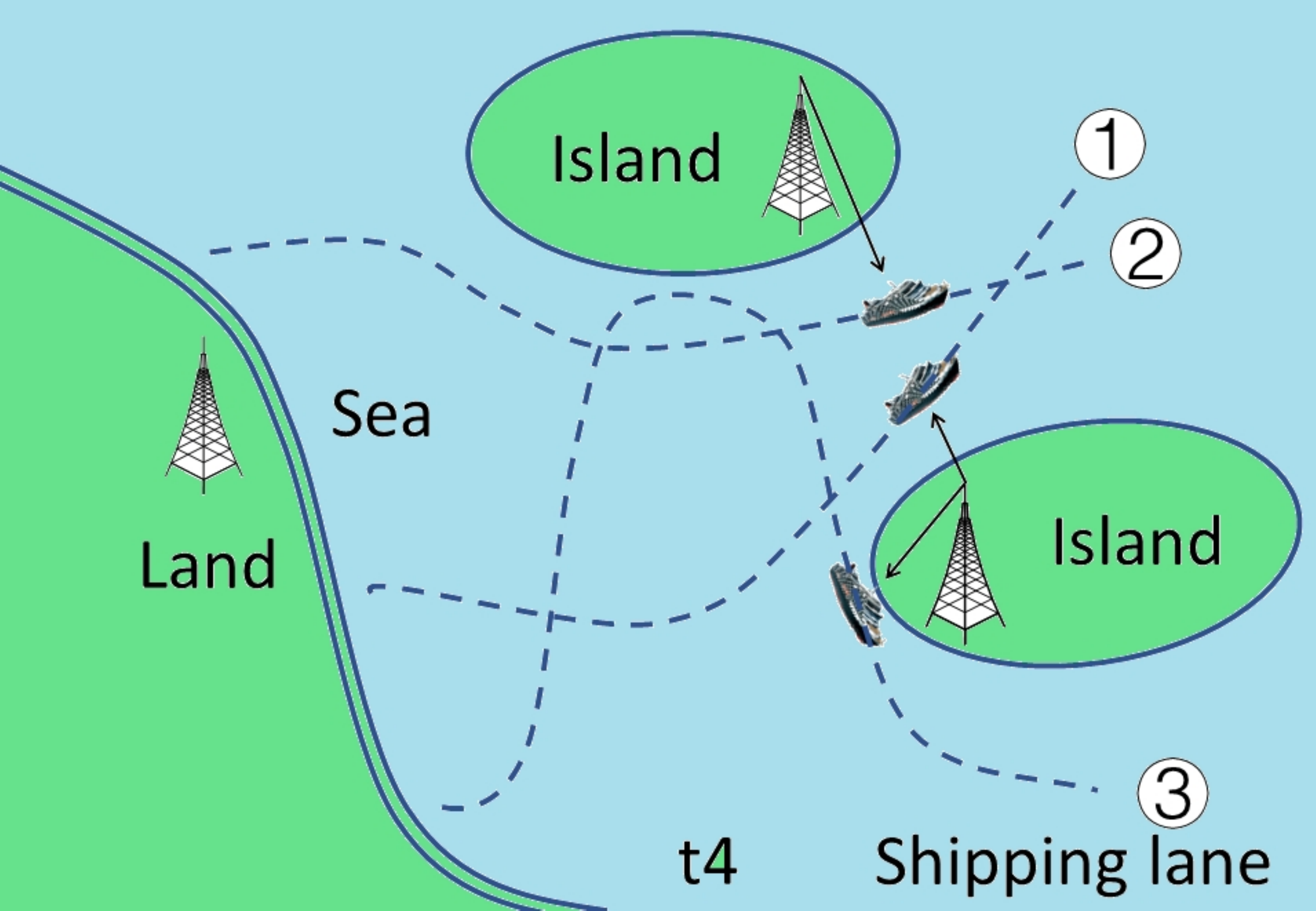}
  }

\vspace{-2mm}

\caption{Illustration of a MCN, where long-term resource allocation is performed based on the shipping lane information. Different from conventional models in terrestrial networks, this model utilizes forward-looking CSIT for global optimization in the temporal dimension. For example, User 2 is not being served at t1 although it is closer to the BS than User 1, because its large-scale channel condition will be even better at t3 and t4 according to its shipping lane.}

\vspace{-4mm}
\end{figure*}

We consider the downlink transmission of a MCN consisting of $J$ onshore BSs and $K$ marine users.
The BS is equipped with $L$ antennas, and each user is equipped with a single antenna.
We assume that orthogonal frequency-division multiple access is used. The total bandwidth available to each BS is $B$, and all subcarriers have identical bandwidth
${{B}_{s}}=\frac{B}{N}$, where $N$ is the number of subcarriers.

We partition the total service duration of the $K$ users into $M$ time slots. The length of each time slot ${\Delta T}$ is chosen so that the large-scale CSIT of the $n^{th}$ $(n=1,2,...,N)$ subcarrier from the $j^{th}$ $(j=1,2,...,J)$ BS to the $k^{th}$ $(k=1,2,...,K)$ user remains constant within each slot $m$ $(m=1,2,...,M)$, which we denote by ${\beta _{k,m,j,n}}$.
Accordingly, we denote the composite channel gain at time instant $t$ in the $m^{th}$ time slot by $\sqrt {{\beta _{k,m,j,n}}} {{\bf{h}}_{k,t,j,n}}$. The small-scale fading vectors are denoted by $\mathbf{h}_{k,t,j,n} \in \mathbb{C}^{L \times 1}$ with the elements following complex Gaussian distribution with standard deviation ${\sigma _s} = 1$, i.e., ${{\bf{h}}_{k,t,j,n}} \sim \mathcal{CN}({\bf{0}}, {{\bf{I}}_{L \times 1}})$. According to the two-ray shore-to-ship propagation model \cite{p2}, the large-scale fading coefficient ${\beta _{k,m,j,n}}$ can be expressed as
\begin{align}
{{\beta }_{k,m,j,n}}={{\left( \frac{\lambda_{j,n} }{4\pi {d}_{k,m,j}} \right)}^{2}}{{\left[ 2\sin \left( \frac{2\pi {{H}_{1,j}}{{H}_{2,k}}}{{\lambda_{j,n}} {d}_{k,m,j}} \right) \right]}^{2}},
\end{align}
where ${\lambda_{j,n} }$ is the wavelength of the $n^{th}$ subcarrier of the $j^{th}$ BS, ${d}_{k,m,j}$ is the distance between the $j^{th}$ BS and the $k^{th}$ user at the ${m^{th}}$ time slot, ${H}_{1,j}$ and ${H}_{2,k}$ represent the antenna heights of the $j^{th}$ BS and the $k^{th}$ user, respectively.
As depicted in Fig. 1, each user sails
according to its designated shipping lane.
With ${d}_{k,m,j}$ known beforehand to the BS based on the shipping lane and timetable,
${{\beta }_{k,m,j,n}}$ is predicted from (1),
and is used for resource allocation over all time slots.
For each user, delay-tolerant information distribution service (for example, video downloading) is assumed, and the amount of data required by the ${k^{th}}$ user is denoted by $C_{k}^{\min }$.


\vspace{-1 mm}
\section{Resource Allocation with Large-Scale CSIT}\label{sec:2}

Our objective is to minimize the average power consumption of all BSs over all time slots by means of joint subcarrier-power allocation, while providing the users with guaranteed QoS.
The optimization problem is formulated as
\begin{subequations}
\begin{align}
& \! \mathop {\min } \limits_{{\bf{P}} } \; { \frac{1}{KM} } \sum\limits_{k = 1}^K \sum\limits_{m = 1}^{M} \sum\limits_{j = 1}^J \sum\limits_{n = 1}^N {P_{k,m,j,n}} \\
& {s.t.} \;\; {P_{k,m,j,n}} \ge 0 \\
& \;\;\;\;\;\;\; \sum\limits_{k = 1}^K \sum\limits_{n = 1}^N {P_{k,m,j,n}}  \le {P_{\max }} \\
& \;\;\;\;\;\;\;  {\Delta T}\cdot\sum\limits_{m = 1}^{M} \sum\limits_{j = 1}^J \sum\limits_{n = 1}^N {{r_{k,m,j,n}}}  \ge C_k^{\min } \\
& \;\;\;\;\;\;\;  {z _{k,m,j,n}} = {\mathop{\rm sign}\nolimits} ({P_{k,m,j,n}}) \\
& \;\;\;\;\;\;\;    \sum\limits_{k = 1}^K {{z _{k,m,j,n}}}  \le 1,
\end{align}
\end{subequations}
where
${P_{k,m,j,n}}$ and ${r_{k,m,j,n}}$ are the transmit power and the expectation of achievable rate from the ${j^{th}}$ BS to the ${k^{th}}$ user on the ${n^{th}}$ subcarrier in the ${m^{th}}$ time slot, respectively, $\mathbf{P}={{\left\{ {{P}_{k,m,j,n}} \right\}}^{K\times M\times J\times N}}$, and ${P_{\max }}$ represents the maximum transmit power of each BS.
By ${z _{k,m,j,n}} \in {\{0,1\}}$ we denote whether the ${n^{th}}$ subcarrier of the ${j^{th}}$ BS is allocated for User $k$ in the ${m^{th}}$ time slot.
In other words,
$\text{sign}\left( x \right)\triangleq \left\{\begin{array}{l}
1,\ x>0 \\
0,\ x=0
\end{array}\right.$,
as ${z _{k,m,j,n}}$ depends only on whether ${{P}_{k,m,j,n}}>0$ or not.
The constraint (2d) is to guarantee that each user's QoS requirement is satisfied. The constraint (2f) indicates that each subcarrier can serve at most one user simultaneously.

The ergodic capacity ${r_{k,m,j,n}}$ in (2d) is expressed as
\begin{align}
\!\!\!\!{r_{k,m,j,n}}
\!=\! \mathbb{E} \left[ {B_s} {{\log }_{2}}\left( \!1+\frac{{{P}_{k,m,j,n}}{{\beta }_{k,m,j,n}}{{\left| {{\bf{h}}_{k,t,j,n}} \right|}^{2}}}{\sigma _{n}^{2}} \!\right)  \right],\!\!\!\!
\end{align}
where ${\sigma _{n}^{2}}$ denotes the noise power of the receiver.

In order to optimize ${P_{k,m,j,n}}$ with only large-scale CSIT, we have to take out the small-scale CSIT ${{\bf{h}}_{k,t,j,n}}$ from (3).
Adopting the random matrix theory \cite{p1001}, we introduce a closed-form approximation for ${r_{k,m,j,n}}$ as
\begin{align}
\begin{array}{l}
{r_{k,m,j,n}}  =  {B_s}\left\{ {{{\log }_2} \left( {{1} + \frac{{{P_{k,m,j,n}}}}{{u_{k,m,j,n}^*\sigma _n^2}}{{{\beta_{k,m,j,n}}}}} \right)} \right.\\
\left. \;\;\;\; + { L\left[ {{{\log }_2}\left( {u_{k,m,j,n}^*} \right) - {{\log }_2}\left( e \right)\left( {1 - \frac{1}{{u_{k,m,j,n}^*}}} \right)} \right]} \right\},
\end{array}
\end{align}
where ${u_{k,m,j,n}^*}$ can be uniquely determined by the following fixed-point equation:
\begin{align}
\!\!\!\!\!\! u_{k,m,j,n}^*  \!=\!  1 \!+\!  {{P_{k,m,j,n}}{{\beta_{k,m,j,n}}}\!{{\left[ \!{L\sigma _n^2 \!+\! \frac{{L{P_{k,m,j,n}}}}{{u_{k,m,j,n}^*}}{{{\beta_{k,m,j,n}}}}} \!\right]}^{ \!\!-\! 1}}}\!\!\!\!.\!\!
\end{align}





There are two major challenges of solving the optimization problem in (2). Firstly, the implicit parameter ${{u_{k,m,j,n}^*}}$ in (4) is coupled with ${{P_{k,m,j,n}}}$ through (5). As a
consequence,
${{r_{k,m,j,n}}}$ is a compound function of ${{P_{k,m,j,n}}}$, making the constraint (2d) intractable.
Secondly, due to the integer constraints (2e) and (2f), the problem is a combinatorial optimization problem.
While the optimal solution can be obtained using integer linear programming solvers or with the brute force method, the computational complexity will be exponential at worst.

To cope with the first challenge, we remove the intractable constraint (2d) and transform ${{r_{k,m,j,n}}}$ into a convex function with the following theorem.

\begin{theorem}
The constraint (2d) is equivalent to
\begin{subequations}
\begin{align}
& {B_s}{\Delta T}\cdot\sum\limits_{m = {1}}^{M}\sum\limits_{j = 1}^J \sum\limits_{n = 1}^N {g\left( {{P_{k,m,j,n}},{\omega_{k,m,j,n}}} \right)}  \ge C_k^{\min } \\
& {\omega _{k,m,j,n}} \geq 0,
\end{align}
\end{subequations}
where
\begin{align}
\begin{array}{l}
\!\!\!\!\!\!g\left( {{P_{k,m,j,n}},{\omega _{k,m,j,n}}} \right) = {\log _2} \left( {{1} + \frac{{{P_{k,m,j,n}}}}{{{e^{{\omega _{k,m,j,n}}}}\sigma _n^2}}{{\beta_{k,m,j,n}}}} \right)\\
\;\;\;\;\;\;\;\;\;\;\;\;\; \;\;\;\; \; { + }\; L {{\log }_2}\left( e \right)\left[ {{\omega _{k,m,j,n}} - 1 + {{e^{{-\omega _{k,m,j,n}}}}}} \right].\!\!
\end{array}
\end{align}
\end{theorem}
\begin{IEEEproof}
See the Appendix.
\end{IEEEproof}




Due to the integer constraints in (2e) and (2f), the problem with constraints (2b), (2c), (2e), (2f) and (6) is still a combinatorial optimization problem.
In order to make it tractable,
we relax the discrete variable ${z _{k,m,j,n}} \in {\{0,1\}}$ to a continuous one ${a _{k,m,j,n}} \in {[0,1]}$. The time-sharing factor ${a _{k,m,j,n}}$ can be considered as the fraction of time that the ${n^{th}}$ subcarrier of the ${j^{th}}$ BS is assigned to User $k$ in the ${m^{th}}$ time slot.
The original problem can be transformed into
\begin{subequations}
\begin{align}
& \! \mathop {\min } \limits_{{\bf{P}},{\bf{A}},{\bf{W}} } \;  { \frac{1}{KM} } \sum\limits_{k = 1}^K \sum\limits_{m = 1}^{M} \sum\limits_{j = 1}^J \sum\limits_{n = 1}^N {a_{k,m,j,n}}{P_{k,m,j,n}}  \\
& {s.t.} \;\;  {\omega _{k,m,j,n}} \geq 0 \\
& \;\;\;\;\;\;\; {P_{k,m,j,n}} \geq 0 \\
& \;\;\;\;\;\;\; \sum\limits_{k = 1}^K \sum\limits_{n = 1}^N {a_{k,m,j,n}}{P_{k,m,j,n}}  \le {P_{\max }} \\
& \;\;\;\;\;\;\;  {B_s}{\Delta T}\cdot \sum\limits_{m = 1}^{M} \sum\limits_{j = 1}^J \sum\limits_{n = 1}^N {a _{k,m,j,n}}{g\left( ... \right)}  \ge C_k^{\min } \\
& \;\;\;\;\;\;\;  {a _{k,m,j,n}} \in [0,1] \\
& \;\;\;\;\;\;\;    \sum\limits_{k = 1}^K {{a _{k,m,j,n}}}  \le 1,
\end{align}
\end{subequations}
where $\mathbf{W}\!=\!{{\left\{ {{\omega}_{k,m,j,n}} \right\}}^{K\!\times\! M\!\times\! J\!\times\! N}}$ and
$\mathbf{A}\!=\!{{\left\{ {{a}_{k,m,j,n}} \right\}}^{K\!\times\! M\!\times\! J\!\times\! N\! \! }}$.


The objective function (8a) is convex, since its Hessian matrix with respect to $\mathbf{P}$ and $\mathbf{A}$ is positive semi-definite. As the transformed constraints (8b)--(8g) are also convex,
(8) is a convex optimization problem, meaning that the primal problem and the dual problem have the same optimal solution. Thus, we propose an iterative algorithm to solve it based on the Lagrangian dual decomposition method.
Particularly, to tackle the problem of recovering the discrete subcarrier allocation result ${\hat a_{{k},{m},{j},{n}}} \in {\{0,1\}}$ from the continuous time-sharing factor ${a _{k,m,j,n}} \in {[0,1]}$, the partial derivative of the Lagrangian function is modified, which is described in detail as follows.

The Lagrangian function is defined as
\begin{align}
\begin{array}{l}
L\left( {{\bf{A}},{\bf{P}},{\bf{W}}, \gamma, \nu, \theta } \right)\\
 = { \frac{1}{KM} } \sum\limits_{k = 1}^K \sum\limits_{m = 1}^{M} \sum\limits_{j = 1}^J \sum\limits_{n = 1}^N {a_{k,m,j,n}}{P_{k,m,j,n}} \\
 - \sum\limits_m^{}\sum\limits_j^{} {{\gamma _{m,j}}\left( {{P_{\max }} - \mathop \sum \limits_k^{}\sum \limits_n^{} {a_{k,m,j,n}}{P_{k,m,j,n}}} \right)} \\
 - \sum\limits_k^{} {{\nu _k}\left(\! {{B_s} \Delta T \sum\limits_m^{}\sum\limits_j^{}\sum\limits_n^{} {{a_{k,m,j,n}}} {{g\left( ... \right)}} \! - \! C_k^{\min }} \!\right)} \!\!\!\!\\
 - \mathop \sum \limits_m^{}\sum\limits_j^{}\sum\limits_n^{} {\theta _{m,j,n}}\left( {1 - \sum\limits_k^{} {{a_{k,m,j,n}}} } \right),
\end{array}
\end{align}
where $\gamma$, $\nu$ and $\theta$ are the Lagrange multipliers of (8d), (8e) and (8g), respectively. The other constraints (8b), (8c) and (8f) can be naturally satisfied in the Karush-Kuhn-Tucker (KKT) conditions \cite{p1003}.
The corresponding Lagrangian dual function is
$y\left( {\gamma ,\nu ,\theta } \right) = \mathop {\inf }\limits_{{\bf{P}},{\bf{A}},{\bf{W}}} \;\left\{ {L\left( {{\bf{P}},{\bf{A}},{\bf{W}},\gamma ,\nu ,\theta } \right)} \right\}$.
The Lagrangian dual problem is given by
\begin{subequations}
\begin{align}
& \! \mathop {\max } \limits_{\gamma ,\nu ,\theta } \; y\left( {\gamma ,\nu ,\theta } \right)   \\
& {s.t.} \;\;  {\gamma ,\nu ,\theta } \geq 0.
\end{align}
\end{subequations}


By solving $\frac{{\partial L\left( { \cdot} \right)}}{{\partial {{P}_{k,m,j,n}}}} = 0$,
we can obtain the optimal transmit power with respect to (10) as
\begin{align}
{\hat P_{k,m,j,n}} = {\left[ {{\nu _k}\frac{{{B_s}\Delta T{{\log }_2}\left( e \right)}}{{{\gamma _{m,j}} + \frac{1}{{KM}}}} - \frac{{{e^{{\omega _{k,m,j,n}}}}\sigma _n^2}}{{{\beta _{k,m,j,n}}}}} \right]^ + },
\end{align}
where $[x]^{+} = {\mathop {\max }(x,0)}$.




A suboptimal subcarrier allocation result can be obtained as follows.
Let us modify $\frac{{\partial L\left( {\cdot} \right)}}{{\partial {{\hat a}_{k,m,j,n}}}}$ with
\begin{align}
\begin{array}{l}
\!\!\!\!\!\!{U_{k,m,j,n}} = \frac{{\partial L\left(  \cdot  \right)}}{{\partial {{\hat a}_{k,m,j,n}}}}  - {\theta _{m,j,n}}\\
\!\!\!\!\!\!=\! { \frac{{{\hat P}_{k,m,j,n}}}{KM} }\!+\!{\gamma _{m,j}}{{\hat P}_{k,m,j,n}} \!-\! {\nu _k}{B_s}\Delta T g\!\left(\! {{{\hat P}_{k,m,j,n}},{\omega _{k,m,j,n}}} \!\right)\!\!\!\!\!\!
\end{array}
\end{align}
which is no longer a function of $\theta_{m,j,n}$.
According to the KKT condition, the $N$ subcarriers are assigned to at most $N$ users with the smallest $N$ ${U_{k,m,j,n}}$s in each time slot, i.e.,
\begin{align}
\!\!{\hat a_{{k^*},{m},{j},{n}}} \!=\! \left\{ \begin{array}{l}
\!\!1,\; \left( {{k^*},{m},{j},{n}} \right) {\rm{ = }}\arg {\mathop {\min }\limits_{k,m,j,n}} \{{{U_{k,m,j,n}}}\} \\
\!\!0,\;{\rm else}
\end{array} \right.\!\!\!\!.\!\!
\end{align}

Finally, the Lagrange multipliers are updated with the subgradient method:
\begin{align}
\!\!\!\gamma _{m,j}^{(i + 1)} \!\!=\!\! {\left[ {\gamma _m^{(i)} \!\!-\!\! \delta _1^{(i)}\!\!\left( {{P_{\max }} \!\!- \!\!\mathop \sum \limits_k^{}\sum \limits_n^{} {a_{k,m,j,n}}{P_{k,m,j,n}}} \right)} \!\!\right]^ {\!\!+} }\!\!,\!\!
\end{align}
\begin{align}
\!\!\!\!\!\!\nu _k^{(i + 1)} \!\!\!\!\!=\!\! {\left[ {\nu _k^{(i)} \!\!\!-\!\! \delta _2^{(i)} \!\!\left(\!\! {{B_s}\Delta T  \! \sum\limits_m^{}\!\sum\limits_j^{}\!\sum\limits_n^{} {\!{a_{k,m,j,n}}} g\left( ... \right) \!\!-\!\! C_k^{\min }} \!\!\right)}\!\! \right]^ {\!\!+} }\!\!\!, \!\!\!\!\!\!
\end{align}
where $\delta _1^{(i)}$ and $\delta _2^{(i)}$ are the step sizes in the $i^{th}$ iteration.

The procedure of the proposed algorithm is described in detail in Algorithm 1.

\begin{algorithm}[h]
\caption{Iterative Long-Term Resource Allocation}
\label{alg:1}
\begin{algorithmic}[1]
\STATE Initially set $\gamma >0$, $\nu >0$, $i=1$, and $I_{{\max }}\in\mathbb{Z}^{+}$.
\STATE Initialize ${\bf{P}}$ with a uniform power distribution, and initialize ${\bf{A}}$ with the subchannel allocation method in \cite{p1005}.
\REPEAT
    \FOR{ $m = 1:M$ }
    \FOR{ $j = 1:J$ }
    \FOR{ $n = 1:N$ }
        \FOR{ $k = 1:K$ }
            \STATE 1) Update ${{\hat P}_{k,m,j,n}}$ according to (11).
            \STATE 2) Calculate ${U_{k,m,j,n}}$ according to (12).
            \STATE 3) Update ${{\hat a}_{k,m,j,n}}$ according to (13).
        \ENDFOR
        \STATE Update ${\gamma}_{m,j}$ according to (14).
    \ENDFOR
    \ENDFOR
    \ENDFOR
    \FOR{ $k = 1:K$ }
        \STATE Update $\nu_k$ according to (15).
    \ENDFOR
     \STATE Update $i = i+1$.
\UNTIL {${\bf{P}}$ converges or $i = I_{{\max }}$.}
\end{algorithmic}
\end{algorithm}

\vspace{-1mm}
\section{Simulation Results}\label{sec:3}

The MCN simulated below consists of 3 onshore BSs and 90 ships sailing within 50 km offshore following their designated lanes.
The system uses a carrier frequency of 1.9 GHz.
The other parameters are set as $L=16$,
${B_s}=2$ MHz, ${H_1}=100$ m, and ${H_2}=10$ m.
The power density of the additive white Gaussian noise is -174 dBm/Hz.

First, we evaluate the effects of service duration $M$ on the average downlink transmit power per BS.
As shown in Fig. 2, the proposed large-scale CSIT aided
scheme outperforms the scheme using instantaneous full CSIT in \cite{p4} when $M$ is larger than a certain threshold, and the gap becomes larger with the increase of $M$.
The reason is that with a larger value of $M$, we actually enlarge the optimization space in the temporal dimension, and therefore utilize more information (i.e., the forward-looking CSIT) for energy-efficient transmissions.


Further, we investigate how the number of subcarriers influences the performance of the proposed scheme.
As can be observed in Fig. 3, the proposed scheme remarkably outperforms the other two schemes in all range of $N$ when $M$ is large enough, and the performance gap between the iterative and equal power allocation schemes becomes wider with the growth of $N$.
The reason is that a larger value of $N$ means larger optimization space,
and the proposed scheme with long-term CSIT can better deal with the
resource competition.

\begin{figure} [htb]
\begin{center}
\vspace*{-4mm}
\includegraphics*[width=9cm]{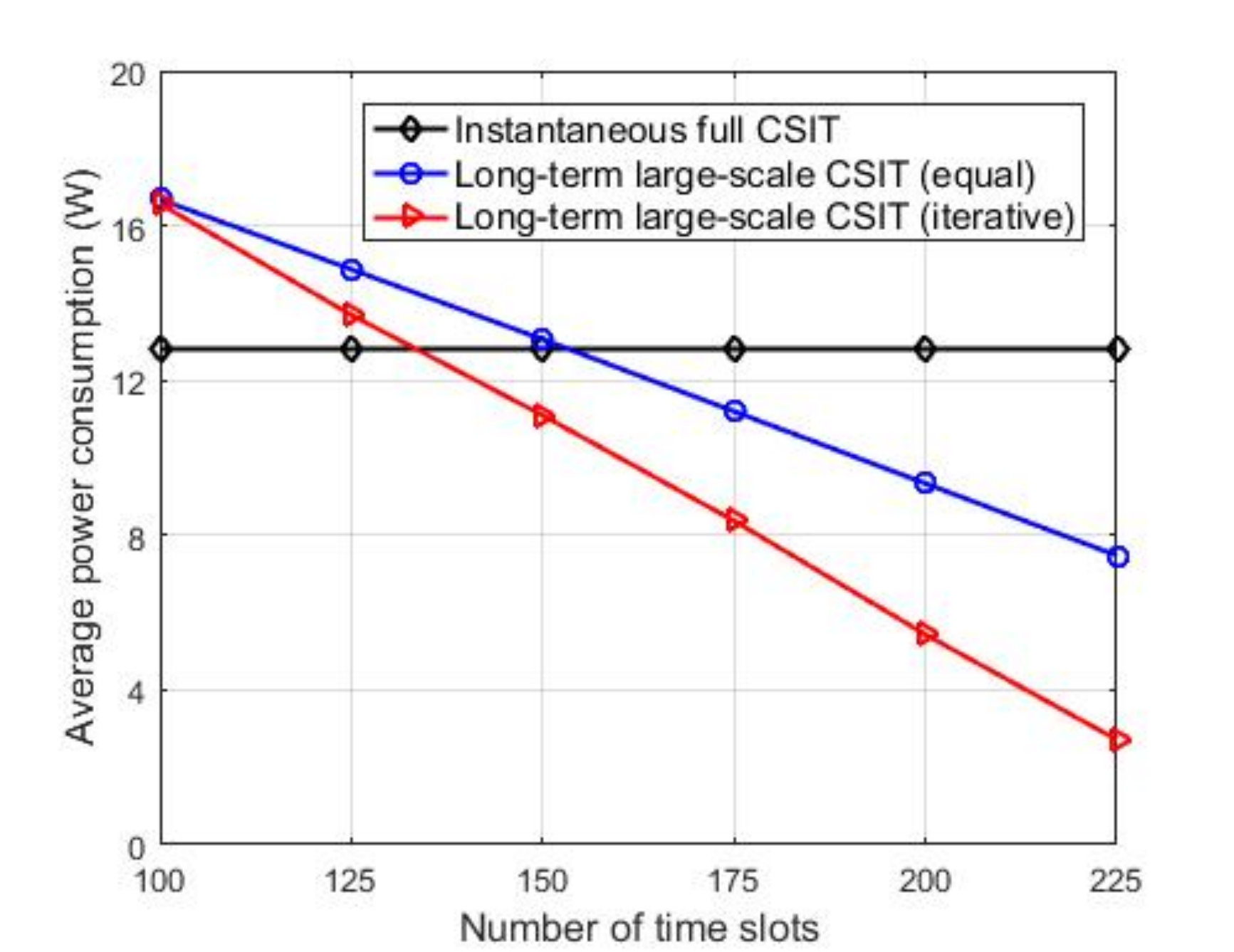} 
\end{center}
\vspace*{-5mm} \caption{Average power consumption with $N=15$ and $P_{\max}=40$ W.}\label{fig:2}  
\vspace*{-4mm}
\end{figure}

\begin{figure} [htb]
\begin{center}
\vspace*{-4mm}
\includegraphics*[width=9cm]{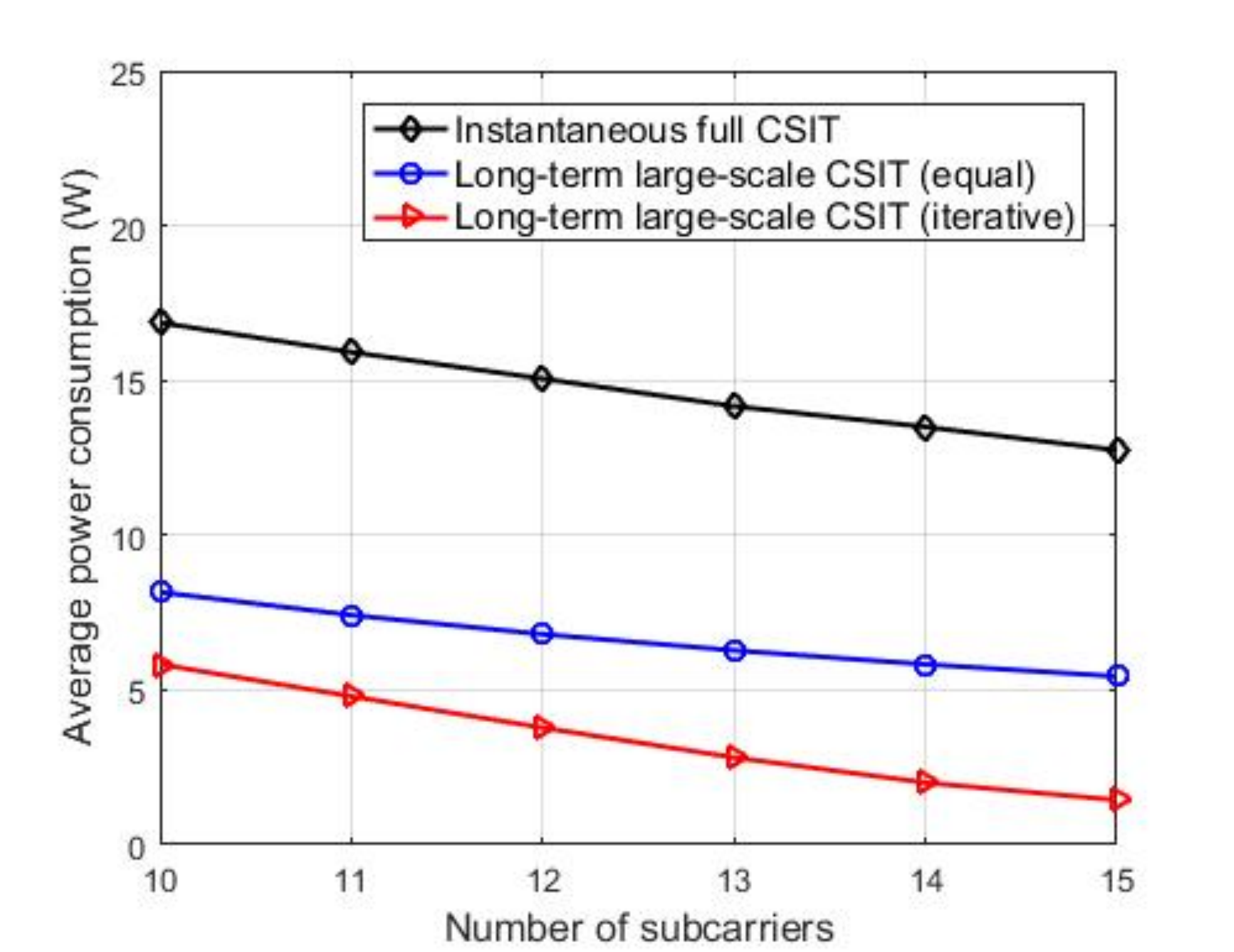} 
\end{center}
\vspace*{-5mm} \caption{Average power consumption with $M\!=\!250$ and $P_{\max}\!=\!40$ W.}\label{fig:2}  
\vspace*{-4mm}
\end{figure}

\vspace*{-1mm}
\section{Conclusions }\label{sec:4}

In this paper, we have focused on enhancing the energy efficiency of MCNs.
By exploiting marine users' position information based on their designated shipping lanes, we have made it possible to estimate the forward-looking large-scale CSIT instead of the complete instantaneous CSIT.
On that basis, we have formulated a long-term joint optimization problem for the allocation of subcarrier and transmit power, to minimize the average power consumption while ensuring the users' QoS requirements.
The problem is non-convex. We have proposed an iterative algorithm to solve it efficiently.
Simulation results have revealed that the proposed scheme can significantly reduce the power consumption.
We have further pointed out that, the performance gain mainly comes from the global optimization over a much larger time span with the shipping lane information,
which implies a brand new way for enhancing the energy efficiency in practical MCNs.


%

\vspace*{-1mm}
\section*{Appendix: Proof of Theorem 1}

Define
\begin{align}
\begin{array}{l}
\!\!\!\!\!\!f\left( {{P_{k,m,j,n}},{u_{k,m,j,n}}} \right) = {\log _2} \left( {{1} + \frac{{{P_{k,m,j,n}}}}{{{u_{k,m,j,n}}\sigma _n^2}}{{\beta_{k,m,j,n}}}} \right)\\
 \;\;\;\;\;\; +  L\left[ {{{\log }_2}\left( {{u_{k,m,j,n}}} \right) - {{\log }_2}\left( e \right)\left( {1 - \frac{1}{{{u_{k,m,j,n}}}}} \right)} \right],\!\!\!\!
\end{array}
\end{align}
and ${{u_{k,m,j,n}^*}}$ in (4) is relaxed into an independent variable ${{u_{k,m,j,n}}}$. As $f\left( {{P_{k,m,j,n}},{u_{k,m,j,n}}} \right)$ is monotonically decreasing with ${u_{k,m,j,n}}$ when $1 \leq {u_{k,m,j,n}} < {{u_{k,m,j,n}^*}}$, and monotonically increasing with ${u_{k,m,j,n}}$ when ${u_{k,m,j,n}} > {{u_{k,m,j,n}^*}}$, the constraint (2d) can be equivalently rewritten as
\begin{subequations}
\begin{align}
&{B_s}{\Delta T}\cdot\sum\limits_{m = {1}}^{M}\sum\limits_{j = 1}^J \sum\limits_{n = 1}^N {f\left( {{P_{k,m,j,n}},{u_{k,m,j,n}}} \right)}  \ge C_k^{\min } \\
& {u _{k,m,j,n}} \geq 1.
\end{align}
\end{subequations}
Define ${{u}_{k,m,j,n}} =  {e^{{\omega _{k,m,j,n}}}}$, and the constraint (17b) can be expressed as (6b). Thus, (2d) can be equivalently transformed into (6). Besides, it can be derived that ${\nabla}_{\omega_{k,m,j,n}}^2 g\left( {{P_{k,m,j,n}},{\omega_{k,m,j,n}}} \right) > 0$. Therefore, $g\left( {{P_{k,m,j,n}},{\omega_{k,m,j,n}}} \right)$ is concave with respect to ${P_{k,m,j,n}}$ and ${\omega _{k,m,j,n}}$.

\end{spacing}

\end{document}